
%
\input harvmac
\baselineskip=12pt
\sequentialequations
\noblackbox

\def\nl{\hfil\break}

\def\npb{{ \sl Nucl. Phys. }}

\def\prd{{ \sl Phys. Rev. }}
\def\prl{{ \sl Phys. Rev. Lett. }}
\def\plb{{ \sl Phys. Lett. }}
\def\rmp{{ \sl Rev. Mod. Phys. }}
\def\IR{\relax{\rm I\kern-.18em R}}
\def\undertext#1{\vtop{\hbox{#1}\kern 1pt \hrule}}
\def\half{{1\over2}}
\def\c#1{{\cal{#1}}}
\def\dirac{\hbox{$\partial$\kern-0.5em\raise0.3ex\hbox{/}}}
\def\dslash{\hbox{$\partial$\kern-0.5em\raise0.3ex\hbox{/}}}
\def\pslash{\hbox{{\it p}\kern-0.5em\raise-0.3ex\hbox{/}}}
\def\gsim{\mathrel{\raise.3ex\hbox{$>$\kern-.75em\lower1ex\hbox{$\sim$}}}}
\def\dk{\int\!\!{d^4\!k\over(2\pi)^4}\,}
\def\dkl{\int_{k^2<\Lambda^2}\!\!{d^4\!k\over(2\pi)^4}\,}
\def\lesssim{\mathrel{\raise.3ex\hbox{$<$\kern-.75em\lower1ex\hbox{$\sim$}}}}
\def\d{\partial}

\def\strivt{s_{triv}^U}
\def\striv{s_{triv}}
\def\st{\tilde S}
\def\nf{N_F}

\Title{\vbox{\baselineskip12pt\hbox{UCLA/92/TEP/34}%
\hbox{UAB--FT--291}}}
{ Non--perturbative evidence for non--decoupling of heavy
fermions}
\centerline{Kenichiro Aoki\footnote{$^1$}
{Work supported in part by the National Science Foundation grant
NSF--89--15286.\nl internet:{\tt~aoki@physics.ucla.edu}}
and Santiago Peris\footnote{$^2$}{Work supported  in part
by the Department of Energy Grant DOE/ER/01545-578 and by the research
project CICYT-AEN90-0028.
\nl bitnet:{\tt~ifteper@ebccuab1}}}
\bigskip\centerline{\it ${}^1$Department of Physics}
\centerline{\it University of California at Los Angeles}
\centerline{\it Los Angeles, CA  90024{\rm --}1547}
\medskip
\centerline{\it ${}^2$Grup de Fisica Teorica}
\centerline{\it Universitat Autonoma de Barcelona}
\centerline{\it E-08193 Bellaterra}
\centerline{\it Barcelona, Spain}
\vskip .3in
\centerline{\bf Abstract}
We investigate, using a $1/N$ expansion, the behavior of a parameter
in the scalar--fermion sector of the standard model
that shows perturbative non--decoupling
as the fermion becomes heavy. This low energy parameter
is related to the $S$ parameter
defined through the $W_3-B$ vacuum polarization tensor.
We obtain the
leading $1/N$ contribution to this
parameter that, if expanded perturbatively,
collapses to its constant one--loop result; remarkably
all the higher--order terms in the series vanish.
Non--perturbatively, however,
we find that as the mass of the fermion approaches the built--in
cutoff scale of the theory ---  the triviality scale --- the parameter
is highly dependent on the implementation of the
cutoff; it is non--universal, and shows non--decoupling.
\bigskip
\Date{10/92}
\newsec{}
When the mass of a particle is generated by a coupling constant,
there are physical effects at low energy that do not vanish as
the particle mass  becomes very heavy.
These so--called non--decoupling effects
\ref\DECOUPLING{T.~Appelquist,
J.~Carazzone,  \prd{\bf11} (1975) 2856 \nl J.C.~Collins,
{\sl ``Renormalization"}, Cambridge University Press (1984), page 223}
are crucial in that
they provide a window into the physics of higher energies
than is currently available.
This is evidenced by the current restrictions placed on the top quark and
Higgs boson masses in the
standard model\ref\WSG{S.~Weinberg,
\prl{\bf 19} (1967) 1264\nl
A.~Salam, in {\sl``Elementary particle theory: Relativistic
groups and analyticity",} Nobel Symp. No. 8, N.~Svartholm ed.,
Almqvist and Wiksell, Stockholm (1969) \nl
S.L.~Glashow, \npb{22} (1961) 579 }
due to precision measurements
\ref\RADIATIVE{For
a review on radiative correction in the standard model, see for instance,
K.~Aoki, Z.~Hioki, R.~Kawabe, M.~Konuma, T.~Muta, {\sl Suppl.
Progr. Theor. Phys.,} {\bf 73} (1982) 1.\nl
For a recent comparison with experiments, see for instance,
P.~Langacker, M.~Luo, A.~Mann, \rmp{\bf64} (1992) 87}.

Non--decoupling effects have also raised an
important issue regarding the attempts to formulate
chiral theories on the lattice
\ref\LATTICE{For recent reviews on the formulation of chiral
fermions on the lattice, see the
proceedings of the International Symposium on Lattice Field Theory,
KEK (1991) and references therein.}.
One of the main problems in this program is
the inevitability of the existence of
the unwanted fermion doublers as required by
the Nielsen--Ninomiya theorem
\ref\NN{H.B.~Nielsen, M.~Ninomiya, \npb{\bf B185} (1981) 20,
{\it ibid.} {\bf B193} (1981) 173}.
In some of the approaches to this problem,
one generates masses of the order of the cutoff scale
of the theory for the unwanted fermions using an effectively
Yukawa--like  coupling.
It has been pointed out that at one loop, this procedure
leaves behind non--decoupling effects, so it
is unlikely to be equivalent to the model without
the unwanted fermions in the low energy theory
\ref\DR{M.~Dugan, L.~Randall, MIT preprint MIT--CTP--2050 (1991)}%
. (Other possible problems have also been pointed out,
some previously \LATTICE
\ref\PROB{
M.J.~Dugan, A.V.~Manohar, \plb{\bf B265} (1991) 137\nl
T.~Banks,\plb{\bf B272} (1991) 75\nl
J.~Smit, \npb{\bf B} (Proc. Suppl.) 4 (1988) 451\nl
M.F.L.~Golterman, D.~Petcher, J.~Smit, \npb{\bf B370} (1992) 52\nl
T.~Banks, A.~Dabholkar, Rutgers preprint RU--92--09 (1992)
\nl S.~Aoki, Tsukuba University preprint, UTHEP--233 (1992)
\nl M.~Golterman, D.~Petcher, lecture given at Rome lattice conference (1992)
\nl and references therein.  }.)

To date, non--decoupling effects have been
studied within perturbation theory mostly to one and, on few occasions, to
two loops.
As the mass of the particle becomes heavier, of course the perturbation
theory becomes less reliable.
It is, therefore, essential to study these issues non--perturbatively and
it is necessary to do so when the mass of the particle is of the order of
the cutoff scale. Such a study will enable us to determine how these
parameters behave outside the perturbative regime and establish the limits
of validity of perturbation theory. Also, there can be, and will be,
important qualitative effects that do not arise within perturbation theory,
as we shall see.

\def\piz{\Pi_{\chi^3}}
\def\yt{y}
\def\yb{y_{\scriptscriptstyle D}}
\def\zp{Z_+}
\def\zz{Z_3}
Let us consider a version of the standard model with spontaneous
breakdown of a {\sl global} SU$(2)_L
\times $U$(1)_Y$ symmetry in which gauge couplings have been turned off.
The effective lagrangian for
the Nambu--Goldstone bosons with heavy fermions integrated out can be
written as
(we use the spacelike signature $(-+++)$ for the metric)
\eqn\nglag{-\c L_\chi= \half\zz\left(\d_\mu\chi^3\right)^2
 + \zp\left|\d_\mu\chi^+\right|^2 +\hbox{\rm interactions}}
In general this lagrangian will be non--local and $\zz,\zp$ will be momentum
dependent functions (in momentum space) amenable to a non--perturbative
calculation in the Yukawa coupling.
Let us define the parameter $\tilde S$ by
\eqn\stdef{\st\equiv-2\pi v^2 \left.{d\over d(p^2)}\zz
(p^2)\right|_{p^2=0}}
where $v$ is the corresponding vacuum expectation value that signals the
spontaneous breakdown of the symmetry. Obviously, $\tilde S$ is ``gauge
invariant" in the sense that it cannot depend on any gauge--fixing
parameters. To one loop
we can compute, for instance, the contribution of a doublet of
massive fermions
to $\tilde S$, and this yields $\tilde S=2 \times {1/( 12 \pi)}$.
It is  mass--independent and in
particular, independent of the amount of the mass splitting.
The two Yukawa couplings cancel out in the definition of $\tilde S$
because  the derivative pulls out two inverse powers of the fermion mass.
So $\tilde S$ shows perturbative non--decoupling, and as a matter of fact, it
counts the number of heavy fermions that have obtained their masses through
the mechanism of spontaneous symmetry breaking.

When we turn the gauge couplings on, the lagrangian of Eq. \nglag\ induces
an interaction between  the gauge bosons when the Nambu--Goldstone bosons
are ``eaten". To lowest order in the gauge couplings $g,g'$ (which are known
to be small) this interaction can be expressed as
\eqn\gaugeit{-\c L_\chi=
\half\zz\left(\d_\mu\chi^3-{v\over2}(gW_\mu^3-g'B_\mu)\right)^2
+\zp\left|\d_\mu\chi^+-{gv\over2}W_\mu^+\right|^2
  +\hbox{\rm other terms}}
where $\zz,\zp$ may be non--perturbative in the Yukawa coupling.
We then discover that $\tilde S$ is the contribution
to lowest order in $g,g'$ of the longitudinal part of the gauge bosons to
the $S$ parameter as defined by Peskin and Takeuchi
\ref\SREF{B.W.~Lynn, M.~Peskin and R.G.~Stuart in ``Physics at LEP", eds. J.
Ellis and R.D. Peccei (1986)\nl
M.~Peskin, T.~Takeuchi, \prl{\bf 65} (1990) 964\nl
D.C.~Kennedy, P.~Langacker \prl{\bf 65} (1990) 2967\nl
G.~Altarelli, R.~Barbieri \plb{\bf 253B} (1991) 161\nl
R.D. Peccei, S.~Peris \prd{\bf D44} (1991) 809.}

\eqn\sdef{S\equiv-\left.{16\pi\over gg'}
        {d\over dp^2}\Pi_{W^3B}\right|_{p^2=0}}
which characterizes the amount of $W^3-B$ mixing and is a measurable
quantity.
For instance, the one--loop contribution of a heavy degenerate doublet to
$S$ and $\tilde S$ are identical, namely $S=\tilde S=1/(6\pi)$
\ref\PP{See the last of the papers in Ref. \SREF}. However, $S$
and $\tilde S$ are not identical in general since, for instance,
 in the case of a
non--degenerate heavy doublet,
$\tilde S$ remains the same but $S$ receives an extra logarithmic
contribution, $S={1\over 6\pi}(1
-Y_L log{m^2_{\scriptscriptstyle U}\over m^2_{\scriptscriptstyle D}})$,
where $Y_L$ is the hypercharge of the left handed
doublet and $m_{U,D}$ are the masses of the up and down--type fermions in
the doublet.

In this paper, we shall study the
non--perturbative behavior of $\tilde S$ when
the Yukawa coupling (or the fermion mass) becomes very large. We choose
$\tilde S$ because it has the same perturbative characteristics as $S$ as
far as non--decoupling is concerned --- which is what we are interested in
studying ---  but allows a much simpler $1/N$--type of non--perturbative
treatment than $S$.
Moreover, it seems quite
reasonable to us that $\tilde S$, being determined by the dynamics of the
symmetry breaking sector, captures the essence of the
non--decoupling phenomena found in the $S$ parameter. After all it is
because of the spontaneous symmetry breakdown that, at least perturbatively,
non--decoupling occurs.

Apart from the phenomenological interest,
non--perturbative aspects  of non--decoupling
effects in renormalizable quantum field theories
are of general importance which should, we believe,
be studied when possible.
In this regard, amongst the presumably trivial theories,
chiral Yukawa theories, such as the
model we are considering, necessarily contain non--decoupling
effects when fermions are massive, making them a natural
setting to study these issues.
Also, these systems have been studied on the lattice extensively
\DR\PROB%
\ref\FLAT{J.~Shigemitsu, \plb{\bf226B} (1989) 364\nl
I-H.~Lee, J.~Shigemitsu, R.~Shrock, \npb{\bf B330} (1990) 225,
\npb{\bf B335} (1990) 265\nl
W.~Bock, A.K.~De, C.~Frick, K.~Jansen, T.~Trappenburg,
\npb{\bf B371}~(1992)~683\nl
S.~Aoki, J.~Shigemitsu, J.~Sloan, \npb{\bf B372} (1992) 361\nl
Proceedings of the International Symposium on Lattice Field Theory,
KEK (1991)\nl and references therein.}
and we hope that our simple calculation might serve
as an useful guide for possible future numerical computations.

\def\zp{Z_+}
\def\zz{Z_3}

\def\mt{m_{\scriptscriptstyle U}}

In a previous study of the $\rho$
parameter\ref\AP{K.~Aoki, S.~Peris, UCLA preprint UCLA/92/TEP/23 (1992)},
it was found that the cutoff effects
in the non--perturbative regime saturated the
perturbative growth with the Yukawa coupling to a constant, making
the behavior milder
than what is naively expected from perturbation theory. Therefore,
an expectation one might harbor in the case of $\tilde S$
is that again cutoff effects diminish the constant value obtained in
perturbation theory, making it vanish as the mass approaches the cutoff.
If this were to happen, this would be very welcome for
the aforementioned problem of decoupling the fermion doublers on the
lattice.  From a different perspective,
$\st$ is independent of the fermion mass to
one loop so that it is somewhat hard to imagine why $\st$
should be sensitive to whether the fermion mass is close
to the cutoff or not. One perhaps expects that while it may
not vanish, it may still be rather insensitive to cutoff effects.
However, we find within the $1/\nf$ expansion that
as the mass approaches the cutoff, the parameter $\st$ does {\it not}
vanish and is cutoff dependent, in other words, it is non--universal.
\newsec{}
The version of the standard model we want to study using the $1/\nf$
expansion has the following lagrangian
\def\ql{q_{\scriptscriptstyle L}}
\def\bl{D_{\scriptscriptstyle L}}
\def\tl{U_{\scriptscriptstyle L}}
\def\tr{U_{\scriptscriptstyle R}}
\eqn\lagrangian{-\c L_\phi=\d_{\mu}\phi^\dagger\d_{\mu}\phi
+\lambda\left(\phi^\dagger\phi-v^2/2\right)^2
 + \overline\ql\dslash\ql + \overline\tr\dslash\tr
+y\left(\overline\ql\phi\tr+\overline\tr\phi^\dagger\ql\right)}
where $\phi$ is in an $\nf$
dimensional irreducible representation of SU($\nf$).
The scalar field develops a vacuum expectation value
$\langle\phi\rangle=(v/\sqrt2\ 0\ 0\ldots0)^T$ that breaks the symmetry of
the lagrangian from U($\nf$) down to U($\nf-1$) and gives mass
to the $U$--fermion. We
define the $\chi^0$ and $\chi^-$ of Eq. \nglag\ as
$\phi=((v+H+i\chi^0)/\sqrt 2, i\chi^-,\ldots)^T$
where $\phi$ has $\nf$ components. There are one massive real
scalar $H$, with tree--level mass $\sqrt{2\lambda}v$, and 2$\nf-1$
Nambu--Goldstone bosons. Within the
fermion sector, $q_L$ and $U_R$ are an $\nf$ and a
1 of SU($\nf$), respectively. We can think of the $q_L$ field as
$\ql\equiv(\tl\bl^1\ \bl^2\ldots\bl^{\nf-1})^T$.

To study the model non--perturbatively, we use the
1/$\nf$ expansion by keeping
$y^2 \nf$, $\lambda \nf$ and $v^2/\nf$ fixed
as we take $\nf$ to infinity. In this
limit, the leading quantum corrections only contribute to the propagator
for the Higgs field, $H$, and the $U$--fermion.
The scalar sector and
the fermion sector can be solved independently. Except for a trivial
shift, $v$ remains unrenormalized so the remaining renormalizations are
only those of $\lambda$ and $y$. We refer the reader to
\ref\EINHORN{M.B.~Einhorn, \npb{\bf B246} (1984) 75}%
\ref\KA{K. Aoki, \prd{\bf D44} (1991) 1547}\AP
for details. Let us only
mention that, to leading order in 1/$\nf$, the $U$--propagator reads
\def\yren{y^2(s_0)}
\def\arb{A_{\scriptscriptstyle R,bare}}
\def\pl{P_L}
\def\pr{P_R}
\def\yb{y_{\scriptscriptstyle bare}}
\def\sbt{s_{\scriptscriptstyle bare}^{\scriptscriptstyle U}}
\eqn\tprop{S_U(p)=\left\{i\pslash\left[\arb(p^2)\pr
+\pl\right]+\yb v\right\}^{-1},\qquad
\arb(p^2)\equiv1-{\yb^2\nf\over2(4\pi)^2}\ln{p^2\over\sbt}\quad.}
where $\pl,\pr$ are projection operators onto the left, right--handed
fields and $\sbt$ denotes a regulator dependent quantity. The Yukawa
coupling is renormalized according to
\eqn\yrenorm{y^2(s_0)={\yb^2\over1-\yb^2\nf/(32\pi^2)\ln s_0/\sbt}}
with an arbitrary renormalization scale $s_0$.
In this renormalization scheme, the renormalized coupling
constant has the physical meaning as the effective
coupling constant at the momentum--squared scale $s_0$, measured,
for instance, through cross sections.
This coupling diverges at a
scale $\strivt\equiv s_0\exp\{32\pi^2/(y^2(s_0)\nf)\}$ which we identify
with the physical cutoff scale in the theory, the triviality scale. This
quantity has the generic form of a non--perturbative effect.
The mass, $\mt$, and the width, $\Gamma_{\scriptscriptstyle U}$,
 of the fermion are determined
from the location of the pole of the full fermion propagator
in the complex plane.
For convenience, we choose the
renormalization scale at the mass scale,
$s_0=\left|\mt-i\Gamma_{\scriptscriptstyle U}/2\right|^2$, in
what follows.
In this convention, since the cross sections need to be finite
at least at the scale of the mass of the fermion,
$\yren$ has to be finite and positive  within the physical region.
In \fig\mplot{The plot of the triviality scale $\striv$,
the mass, $\mt$,
and the width, $\Gamma_{\scriptscriptstyle U}$, against
the renormalized coupling constant $y^2(s_0)$.}
we show the fermion mass and width as
well as the triviality scale as a function of the Yukawa coupling $y^2(s_0)$
evaluated at a scale $s_0$.
The mass of the fermion is smaller than $5.0\sqrt{v^2/\nf}$
when the coupling constant $\yren$ is positive and finite.

This 1/$\nf$ generalization of the standard model is largely dictated by
simplicity. For instance, it would be of interest to study also the case
where custodial symmetry is unbroken,
perhaps using the large--$N$ limit of
\ref\EG{M.B.~Einhorn, G.~Goldberg, \prl{\bf57 }(1986) 2115}. However, the
model in this case seems substantially more complicated.

\def\ar{A_{\scriptscriptstyle R}}
\def\hmt{\hat\mt}
\def\ark{\ar\!(k^2)}
\def\arksq{\ar^2\!(k^2)}
The leading order corrections to the two point function
of the neutral Nambu--Goldstone boson, $\piz$, arise
from the class of one--particle irreducible
graphs in \fig\figse{The class
of one--particle irreducible
graphs contributing to the propagator of the neutral
Nambu--Goldstone boson. Dashed and solid
lines represent Nambu--Goldstone bosons
and fermions, respectively.}
and is of the order $\c O(1/\nf)$.
The contribution of a fermion multiplet to
$\piz$ may be computed as \AP
\eqn\pizeq{ \piz(p^2)=2\yt^2\!(s_0)\dk{\ar((k+p)^2)k(k+p)+\hmt^2\over
\left[\ar(k^2)k^2+\hmt^2\right]
\left[\ar((k+p)^2)(k+p)^2+\hmt^2\right] }}
where $\hmt^2\equiv\yt^2\!(s_0)v^2/2$ and
$\ar(s)\equiv1-\yt^2\!(s_0)\nf/(32\pi^2)\ln s/s_0$.
The wave function renormalization factor $\zz$ in \nglag\
is related to this contribution as
\eqn\zzeq{\zz(p^2)=1-{d\over dp^2}\piz(p^2)}
Using this relation and the definition of $\st$ in \stdef,
we obtain the following expression (in euclidean space)
after some algebra:
\eqn\sans{\st={16\pi\hmt^4\over3}
\dkl{\c N\over k^2(\ar(k^2)\,k^2+\hmt^2)^5}}
with
\def\aly{\alpha_y}
\eqn\ndef{\c N\equiv\aly\,\,
(k^2)^2\left[\arksq-3\aly\ark+3\aly^2\right]
  +\hmt^2k^2\left[3\arksq-7\aly\ark+6\aly^2\right]+\aly\hmt^4}
where we used the shorthand $\aly\equiv\yt^2\!(s_0)\nf/(32\pi^2)$.
In the above expression, there is a pole in the integrand
above the triviality scale so that the integral is ill--defined
unless we restrict the integration region.
The pole is always larger than the triviality
scale so that we cutoff the integral at a scale $\Lambda^2$ below
$\striv$, which is consistent with the existence of the intrinsic
cutoff scale $\striv$ in the theory. This is how the physical cutoff comes
to play the active role that one naturally expects and that is always
missed in any perturbative treatment.
The integral \sans\ may be computed after some work to be
\def\xl{x_{\scriptscriptstyle \Lambda}}
\def\arl{\ar\!(\Lambda^2)}
\eqn\sanal{\st={1\over12\pi}
  \left[1+{\xl^2\aly(-2\arl+3\aly)+4\xl(-\arl+\aly)-1\over
  (\arl\xl+1)^4}\right]\quad\hbox{where }\xl\equiv{\Lambda^2\over\hmt^2}}

If we expand this expression  for $\st$ in powers
of the coupling constant as we would in perturbation theory,
the need to restrict the integration region disappears.
The truly remarkable fact regarding this parameter in this case
is that to {\it all } orders in perturbation theory,
this parameter $\st$ is $1/(12\pi)$ and is independent
of the Yukawa coupling, or equivalently
the fermion mass, to leading order in the $1/\nf$ expansion;
in other words, all the higher order terms in the
expansion for $\st$ in \sans\ surprisingly cancel.
In fact, it is clear from \sanal\ that $\st$ reduces
to its constant value in the limit cutoff goes to infinity.
The above expressions for $\st$ in \sans\ or \sanal\
 include contributions
from one--particle irreducible graphs of arbitrary high order
(cf. \figse) and these contributions are ultimately crucial, so that this
is not a trivial fact.
The dependence of $\st$ on the mass of the fermion,
then, comes solely from the necessity of imposing the
cutoff in the theory, which makes this parameter an ideal
setting for investigating the physical effects of
the triviality cutoff.

We may compute the parameter numerically and our results
are plotted in \fig\syplot{$12\pi\st$ plotted against the renormalized
coupling constant $\yt^2\!(s_0)$
for the values of $\Lambda/\sqrt{\striv}=0.1,0.5$
and $0.8$, which we call ``cutoff" in the plot.} and
\fig\smplot{$12\pi\st$ plotted against $\mt/v$
for the values of $\Lambda/\sqrt{\striv}=0.1,0.5$
and $0.8$, which we call ``cutoff" in the plot.} against the
renormalized coupling constant $\yt^2\!(s_0)$ and the mass
of the fermion $\mt$, respectively for a few cutoff values,
$\Lambda/\sqrt{\striv}=0.1,0.5$ and $0.8$. As the Yukawa coupling
grows, the cutoff $\Lambda$ decreases and eventually the physical fermion
mass would
be larger than the cutoff.
We have plotted only the region where the fermion
mass is smaller than the corresponding cutoff, $\Lambda$.
All calculations agree in the perturbative regime.
As the mass approaches the cutoff scale, the
results depend on the cutoff scale and deviate from
the perturbative result.
As the mass increases to $\mt=3.12v$,
the $\st$ parameter computed with $\Lambda/\striv=0.8$
differs 1\%\ from the perturbative result, at which point,
$\sqrt{\striv}=90v$, $\Gamma_{\scriptscriptstyle U}/\mt=0.51$
and $\yt^2\!(s_0)=23.5$.
The maximum mass of the theory in the large--$\nf$ limit
is 3.52$v$ so that the deviations from the perturbative
result are appreciable only when the mass is close to
its maximum value.
As we can see, the contribution to $\st$ does not vanish
within the physical region defined by $\mt< \Lambda$, although there is an
apparent decreasing trend at large couplings that is stronger for low values
of the cutoff. If there is a way to make sense of the region $\mt>\Lambda$
in some framework, whether $\st$ can vanish in this region
might deserve some further investigation.

In closing, we point out that this contribution to $\st$ can
be understood as the effect of operators of dimension eight or higher
in the effective scalar theory.
At dimension eight, there is effectively only one operator,
$\c O \sim \phi^{\dagger}  D_{\mu} D_{\nu}
\phi \  \phi^{\dagger} D^{\mu} D^{\nu} \phi$ that contributes
to $\st$.
The first constant term in \sanal\ is generated by an
operator like $\c O/v^4$ and the cutoff dependent terms
are generated by $\c O/\Lambda^4$, in both cases,
up to higher dimension operators.
The former does not fall off with the cutoff and is
a perturbatively relevant, however a cutoff {\it independent}
contribution.
The latter is a cutoff dependent but a perturbatively
irrelevant contribution. The sole reason this
term is not negligible is because the cutoff scale cannot
be taken to infinity since it needs to be smaller than
the triviality scale.
\bigskip\noindent{\it Acknowledgments:}
K.A. would like to thank Roberto Peccei
for discussions and Hidenori Sonoda for his insightful
discussions and suggestions. K.A. also would like
to acknowledge the hospitality of
the Aspen Center for Physics, where some of this work
was conducted. S.P. would like to thank E. Masso and P. Hernandez for
conversations. Most of the work of S.P. was done while he still was at the
Ohio State University and he wishes to thank the members of the Physics
Department of this institution for their warm hospitality.
 \listrefs
 \listfigs
\end